\documentclass[letterpaper, 10 pt, conference]{ieeeconf}

\IEEEoverridecommandlockouts                              
\usepackage{amsmath,amsfonts,amssymb}
\newcommand\inv[1]{#1\raisebox{1.15ex}{$\scriptscriptstyle-\!1$}}
\usepackage{graphics} 
\usepackage{epsfig,color} 
\usepackage{times} 
\usepackage{amssymb}
\usepackage[
  separate-uncertainty = true,
  multi-part-units = repeat
]{siunitx}
\usepackage{amsfonts}
\usepackage{amsthm}
\usepackage{graphicx}
\graphicspath{ {./images/} }
\usepackage{graphicx}
\usepackage{subfig}
\usepackage[ruled,vlined]{algorithm2e}
\usepackage{footnote}
\newtheorem{assumption}{Assumption}
\newtheorem{remark}{Remark}
\newtheorem{theorem}{Theorem}

\title{\LARGE \bf
Singular Perturbation-based Reinforcement Learning of Two-Point Boundary Optimal Control Systems
}

\author{Vasanth Reddy, Hoda Eldardiry and Almuatazbellah Boker% <-this % stops a space
\thanks{Vasanth Reddy and Hoda Eldardiry are with Department of Computer Science,
        Virginia Tech, Blacksburg, VA, USA.
        {\tt\small emails: vasanth2608@vt.edu and hdardiry@vt.edu}}%
\thanks{Almuatazbellah Boker is with Bradley Department of Electrical and Computer Engineering, Virginia Tech, Blacksburg, VA, USA.
        {\tt\small email: boker@vt.edu}}%
}

\begin{document}

\maketitle
\thispagestyle{empty}
\pagestyle{empty}
\begin{abstract}
This work presents a technique for learning systems, where the learning process is guided by knowledge of the physics of the system. In particular, we solve the problem of the two-point boundary optimal control problem of linear time-varying systems with unknown model dynamics using reinforcement learning. Borrowing techniques from singular perturbation theory, we transform the time-varying optimal control problem into a couple of time-invariant subproblems. This allows the utilization of an off-policy iteration method to learn the controller gains. We show that the performance of the learning-based controller approximates that of the model-based optimal controller and the accuracy of the approximation improves as the time horizon of the control problem increases. Finally, we provide a simulation example to verify the results of the paper.\end{abstract}

\begin{keywords}
optimal control; singular perturbation; reinforcement learning
\end{keywords}

\section{INTRODUCTION}
Control and analysis of time varying (dynamic) systems has been extensively investigated over the last decade. Applications such as rocket landing~\cite{c5, c6}, energy conservation in electronic circuits~\cite{c7, c8} can be categorized as finite-horizon time-varying systems.
Solving this category of problems requires solving the time varying Riccati equation for linear time-variant systems or time-varying Hamiltonaian Jacobi equation for nonlinear systems. Solving time-variant equation is bit complicated and tough to solve than compared to the time-invariant equations. The work in ~\cite{c9} illustrates that the two value boundary problems exhibits two-time scale phenomenon even if the original system is not singularly perturbed. In this case, the work~\cite{c9} shows that it is possible to approximate the original time-varying system into two time-invariant systems. One is initial boundary problem which stabilizes the original system in forward time and the other is the final boundary system which stabilized the terminal layer in reverse time. As the perturbation parameter, which is related to the time-horizon of the control problem, decreases the accuracy of the approximation increases. A number of research papers have shown the use of the above approximation in nonlinear, fixed end point control problems~\cite{c10}, nonlinear problems with singular arcs~\cite{c11} and in optimal control of quasi linear systems~\cite{c12}. In the recent times, this approximation is well used to study the turnpike properties in wave equations~\cite{c13} and in semi-linear control~\cite{c14}. \par
The approximation solution for the two value boundary is provided assuming that the dynamics of the model is known. However, in the real time applications the dynamics of model may not be known, or suffer from modeling uncertainties, and in that case it is difficult to solve the initial and final boundary problems. In the past, the research work~\cite{c15} introduced adaptive controls laws through which we can learn the control of the model from the input-output data. In the recent years the gain in momentum of Reinforcement Learning~\cite{c16} have opened up the different possibilities of learning the controller. Adaptive Dynamic Programming is one of the first learning techniques in Reinforcement Learning. It uses an iteration method proposed in previous work~\cite{c17} with a change in variables to find the optimal control gain. Learning methods such as Q-learning~\cite{c18}, Actor-Critic~\cite{c19} have been effective in learning the controller for continuous time systems when the model dynamics are unknown. \par
The main aim of this paper is to learn the controller for the finite horizon time varying system when boundary conditions are given.  In the past, research work has been carried out in learning the optimal control when the system dynamics are unknown using adaptive dynamic programming~\cite{c17}, actor-critic~\cite{c19} and Q-learning~\cite{c18} but they assume that the system is time invariant. Extending this work to the time varying systems, later work~\cite{c20} proposed an data-driven approach and other work~\cite{c21, c22} used policy iteration method to find the controllers but they are restricted to only discrete time-varying systems. Previous work~\cite{c23} investigated continuous time periodic systems which is an important applications of the time-varying systems. It used adaptive programming to learn the controller and off-policy value iteration~\cite{c24} but considered the infinite horizon without any boundary conditions. Limited work has been done and proportionately less results are available for finite horizon time varying systems. More recent work~\cite{c25} proposed dual loop iterative algorithm but they considered the fact that the state transition matrix is unknown. \par
In order to address the above problems we adopt the idea of solving the finite horizon time varying problems with given boundary conditions from seminal work~\cite{c3}. The idea is: When the time period in which the cost function needs to be optimized is large, the time-varying system exhibits the two-time scale property, where the system dynamics evolve at faster time scale relative to the time-scale of the control effort. Then \cite{c3} shows that the system can be reduced into two time-invariant problems. The final original state thus can be approximated by superimposing the solutions of initial and final layer problems. We further use the idea of offline learning~\cite{c26} to estimate the control gain of two boundary problems. Thus we reduce the complexity of original system into two simple problems and then learning the two individual controllers. We empirically show that the controller learned from two individual is nearly approximate to that of the original controller. We also show that the accuracy of the state and controller increases as the time interval to optimize the cost function increases. We demonstrate the above point by taking up a linear time-variant system example. \par
In summary, the key contributions of our proposed model are twofold: 
\begin{enumerate}
    \item We provide a solution for the two-boundary optimal control problem for linear time-varying systems that does not require knowledge of system model.
    \item We introduce a reinforcement learning framework that is based on an insight from the physical dynamics. This framework is simple to implement and results in a suboptimal solution, which converges to the optimal one as the control time interval gets large.%This is a step towards making machine learning algorithms more explainable.  
\end{enumerate}
The rest of this paper is organized as follows. Section~\ref{problem formulation} illustrates the system setup and problem formulation. Section~\ref{Main result} describes the two-time scale reduction of original problem. Section~\ref{learning} outlines the learning algorithm to estimate the control gains of the system. Section~\ref{simulation} presents the numerical example and illustrates the simulations. 
\section{Problem Formulation} \label{problem formulation}
Consider the linear time-varying system expressed by the differential equation: 
\begin{equation} \label{systemeq}
    \frac{dx}{dt}= A(t)x(t) + B(t)u(t)
\end{equation}
where $x(t) \in \mathbb{R}^n$, $u(t) \in \mathbb{R}^m$, $A(t) \in \mathbb{R}^{n \times n}$ and $B(t) \in \mathbb{R}^{n \times m}$ are the system states, control input, state matrix and input matrix, respectively. The matrices $A(t)$ and $B(t)$ are smooth functions of time $\forall\; t \in [0, T]$ and can be unknown.
The control objective is to design $u(t) \in \mathbb{R}^m$ to drive the system states $x(t)$ from the initial state $x(0) = x_0$ to the final state $x(T) = x_T$ within a time period $T$ and meanwhile minimize the objective function
\begin{equation}\label{costfunction}
    J = \int_{0}^{T} x^\top(t)Q(t)x(t) + u^\top(t)R(t)u(t) \hspace{2mm} dt
\end{equation}
where, $Q(t) = Q^\top(t) \succeq 0$ and $R(t) \succ 0\;\;\forall\; t \in [0, T].$ The matrices $Q(t)$ and $R(t$) are assumed to be smooth functions of time.
We further have the following assumptions:
\begin{assumption}\label{as1}
The pair $(A(t), B(t))\;\;\forall\; t \in [0, T]$ is controllable and the pair $(A(t), \sqrt{Q(t}) )\;\;\forall\; t \in [0, T]$ is observable.
\end{assumption}
\begin{assumption}\label{as2}
The time $T$ needed to accomplish the control objective is long compared to the system dynamics. 
\end{assumption}
Assumption~\ref{as1} is a standard assumption according to the theory of optimal control~\cite{c101}. Assumption~\ref{as2} implies that the system is slowly varying relative to the control objective. This will be explained further in the next section.  

To solve the optimal control problem, we define the Hamiltonaian function for the system (\ref{systemeq}) as~\cite{c101}: 
\begin{multline} \label{hamiltonian}
\mathcal{H} = x^\top(t)Q(t)x(t) + u^\top(t)R(t)u(t) + \\
p^\top(t)(A(t)x(t) + B(t)u(t))
\end{multline}
where, $p(t) \in \mathbb{R}^n$ is the co-state equation that satisfies: 
\begin{equation} \label{costate}
    \dot{p}(t) = -\nabla_x \mathcal{H} = -Q(t)x(t) - A^\top(t)p(t).
\end{equation}
The control law $u(t)$ for the dynamic system (\ref{systemeq}) that minimizes the objective function (\ref{costfunction}) is given by $\nabla_u \mathcal{H} = 0$: 
\begin{equation} \label{control}
    u(t) =  -\inv{R}(t)B^\top(t)p(t).
\end{equation}
Our objective is to learn the optimal controller $u(t)$ given by \eqref{control} for the system \eqref{systemeq} without the need to know the system matrices $A(t)$ and $B(t)$.  
%\section{Main Results} \label{Main result}
\section{Singular Perturbation-based Design} \label{Main result}
It is known that for control problems that are formulated over a finite time interval $t\in[0,T]$, the closed-loop system dynamics behave in two time scales~\cite{c102}. The separation of time scales is dependent on the length of the time interval $T$. In this work, we leverage this phenomenon by considering the case when the time needed to transfer the system state from one instant to another is small compared with $T$. This allows the system to be modeled in a singularly perturbed form. Accordingly, previous work~\cite{c102} has shown that it is possible to achieve a sub-optimal solution to the control problem when $T$ is sufficiently long and matrices $A(t)$ and $B(t)$ are known. In the next sections, we show that it is possible to achieve a similar result without requiring the system matrices to be known thanks to employing a reinforcement learning algorithm.
Towards this goal, we start by setting up the singular perturbation model of the system. Normalizing the time period $0$ to $T$ to the interval $[0, 1]$ by introducing the scaled time 
\begin{equation}
    \tau = \frac{t}{T} \label{tau}
\end{equation} and defining
\begin{equation}
 \varepsilon =1/T   
\end{equation}
and then reconstructing~(\ref{systemeq}),~(\ref{costfunction}),~\eqref{costate}, and~(\ref{control}) to obtain
\begin{align}
    \varepsilon\begin{bmatrix}
    \frac{dx}{d\tau}\\
    \frac{dp}{d\tau}\end{bmatrix}=&
    \begin{bmatrix}
    A(\tau)&& 0\\-Q(\tau) &&-A^T(\tau)
    \end{bmatrix}\begin{bmatrix}
    x\\p
    \end{bmatrix}+\begin{bmatrix}
    B(\tau)\\0
    \end{bmatrix}u \label{systemeq1},\\
    J =& T*\int_{0}^{1} x^\top(\tau)Q(\tau)x(t) + u^\top(\tau)R(\tau)u(\tau) \hspace{2mm} dt,\label{costfunction1}\\
    u(\tau) =& -\inv{R}(\tau)B^\top(\tau) p(\tau), \label{cont}
\end{align} 
where  $x(0) = x_0$ and $x(1) = x_T$.
For the system described by~(\ref{systemeq1}) to be in standard singular perturbation form, we make the following assumption:
\begin{assumption} 
The eigenvalues of the Hamiltonian matrix
\begin{equation}
    M_H\triangleq\begin{bmatrix} A(\tau) &&-B(\tau)R^{-1}(\tau)B^\top(\tau)\\-Q(\tau)&&-A^\top(\tau)
    \end{bmatrix}
\end{equation}
lie off the imaginary axis $\forall\; t\in[0,1]$.
\end{assumption}  
This assumption implies that $M_H$, which is the closed-loop matrix after substituting \eqref{cont} in \eqref{systemeq1}, is nonsingular $\forall\; t\in[0,1]$.

We are going next to decouple the dynamics of \eqref{systemeq1}. To accomplish this task, we use the transformation~\cite{c102}
\begin{equation} \label{trans1}
\left[\begin{array}{l}
x \\
p
\end{array}\right]=\left[\begin{array}{cc}
I & I \\
P_a(\tau, \varepsilon) & P_b(\tau, \varepsilon)
\end{array}\right]\left[\begin{array}{l}
x_a \\
x_b
\end{array}\right].
\end{equation}
It is shown in [Lemma 2.3,~\cite{c102}] that, under the assumptions of this paper, the transformation \eqref{trans1} is nonsingular for sufficiently small $\varepsilon.$
Accordingly, using \eqref{trans1}, system \eqref{systemeq1} with \eqref{cont} can be transformed into 
\begin{align} 
    \varepsilon \frac{d}{d\tau}x_a =& A(\tau)x_a + B(\tau)u_a(\tau), \label{l} \\
    \varepsilon \frac{d}{d\tau}x_b =& A(\tau)x_b + B(\tau)u_b(\tau)\label{r},
\end{align}
where
\begin{align}
    u_a =&-\inv{R}(\tau)B^\top(\tau)P_a(\tau, \varepsilon),\\
    u_b =&-\inv{R}(\tau)B^\top(\tau)P_b(\tau, \varepsilon),
\end{align}
and $P_a(\tau, \varepsilon) \geq 0$ and $P_b(\tau, \varepsilon) \leq 0$ are the roots of the differential Riccati equation
\begin{equation} \label{diffric1} 
\begin{aligned}
\varepsilon \dot{P} = &-A^\top(\tau)P - PA(\tau)\\
&+ PB(\tau)\inv{R}(\tau)B^\top(\tau)P - Q(\tau). 
\end{aligned}
\end{equation}
We transform next the singular perturbation system \eqref{l}-\eqref{r} into the boundary layer system by considering the time-scale change
\begin{equation}\label{ch}
 \gamma = \frac{\tau}{\varepsilon},\quad \beta = \frac{1-\tau}{\varepsilon}.   
\end{equation}
In view of \eqref{ch}, the boundary layer system is obtained by setting $\varepsilon \to 0$ in \eqref{l}-\eqref{diffric1}. This leads to the \textit{initial regulator problem}
\begin{equation} \label{iniregulator}
   \frac{d}{d\gamma}x_a = A(0)x_a + B(0)u_a,
\end{equation}
with the feedback controller $u_a$ in the form
\begin{equation} \label{u_a}
    u_a(\gamma) \triangleq -K_ax_a(\gamma)= -\inv{R}(0)B^\top(0)P_a(0)x_a(\gamma),
\end{equation}
which minimizes the cost function
\begin{equation} \label{j_a}
    J(x_a, u_a) = \int_{0}^{\infty} x_a^\top Q(0)x_a+ u_a^\top R(0) u_a \hspace{1mm} d\gamma,
\end{equation}
and the \textit{terminal regulator problem} \begin{equation} \label{finregulator}
   \frac{d}{d\beta}x_b = -A(1)x_b- B(1)u_b,
\end{equation}
with the feedback controller 
\begin{equation} \label{u_b}
    u_b(\beta)\triangleq -K_bx_b(\beta) = -\inv{R}(1)B^\top(1)P_b(1)x_b(\beta),
\end{equation}
which minimizes the cost function
\begin{equation} \label{j_b}
    J(x_b, u_b) = \int_{0}^{\infty} x_b^\top Q(1)x_b + u^\top_b R(1)u_b \hspace{1mm} d\beta.
\end{equation}
Based on this setup, we have the following theorem.
\begin{theorem} [Theorem 2.1 in \cite{c102}]\label{theorem}
Under Assumptions 1-3, there exists $\varepsilon^*>0$ such that for all $\varepsilon\in(0,\varepsilon^*]$ the solution $x(\tau), p(\tau)$ of \eqref{systemeq1} and the optimal control $u(\tau)$ satisfy
\begin{align}
    x(\tau)=&x_a(\gamma)+x_b(\beta)+\mathcal{O}(\varepsilon),\label{x}\\ 
    p(\tau)=&P_a(0)x_a(\gamma)+P_b(1)x_b(\beta)+\mathcal{O}(\varepsilon), \label{p}\\
    u(\tau)=&-K_ax_a(\gamma)-K_bx_b(\beta)+\mathcal{O}(\varepsilon). \label{c}
    \end{align}
\end{theorem}
Theorem~\ref{theorem} implies that if the initial and terminal regulator problems, which are now problems for linear time-invariant systems, are solved then the solution will approximate that of the original control problem \eqref{systemeq}-\eqref{control} for sufficiently small $\varepsilon$. 
In the next section, we will rely on this result and develop a reinforcement learning algorithm to solve the initial and terminal regulator problems \eqref{iniregulator}-\eqref{j_b} without the need for $A(t)$ and $B(t)$ to be known. 
\section{Reinforcement Learning} \label{learning}
In this section, we follow a reinforcement learning approach to learn the initial and terminal regulator problems \eqref{iniregulator}-\eqref{j_b}. In particular, we learn the controllers using an offline policy.
\begin{remark} \label{remark1}
In view of Theorem \ref{theorem}, we use the original state $x(t)$ for the offline learning procedure. It should be noted that, according to \eqref{tau} and \eqref{ch}, the time-scales $\gamma$ and $\beta$ are the forward and reverse times of $t$, respectively.
\end{remark}
(i) Initial Regulator problem: The objective is to learn the feedback control gain $K_a\triangleq  \inv{R}(0)B^\top(0)P_a(0)$ for the system described in~\eqref{iniregulator} without knowing the system dynamics. 
In view of Remark~\eqref{remark1}, learning is done using the online measurement data of system states $x$ and control $u_a$ such that control $u_a = -K_ax$ optimizes the cost function described in~\eqref{j_a}.
Note that $P_a(0)$ is the solution to the algebraic Riccati equation: 
\begin{equation} \label{are}
\begin{aligned}
    A^\top(0)P_a(0) + P_a(0)A(0) - P_a(0)B(0)&\inv{R}(0)B^\top(0)P_a(0)\\
    &+ Q(0) = 0.
\end{aligned}
\end{equation}
The optimal feedback gain is given by
$K^*_a = -\inv{R}(0)B^\top(0)P_a $. The optimal values are then found using the Kleimanns algorithm~\cite{c100} as follows. 
\begin{enumerate}
    \item Solve for $\Tilde{P}^k$ of the lyapunov equation
    \begin{equation} \label{ly}
    \begin{aligned}
        A_k^\top(0) \Tilde{P}_a^k+&\Tilde{P}_a^k A_k(0) +Q(0)\\
        &+\Tilde{P}^k_a B(0) \inv{R}(0)B^\top(0) \Tilde{P}^k_a=0.
    \end{aligned}
    \end{equation}
    \item Update the feedback gain:
    \begin{equation} \label{rcgain}
        \Tilde{K}^{k+1}_a = \inv{R}(0)B^\top(0)\Tilde{P}^k_a,
    \end{equation}
\end{enumerate}
where $A_k(0) = A(0) - B(0)K^k_a$. The matrix $P_a(0)^k$ and the gain $K_a^{k+1}$ can be learned iteratively following the above steps but only when the model dynamics are known. As we assumed that the model dynamics are unknown, we follow a few steps to eliminate the use of model dynamics $A(0)$ and $B(0)$ in learning the controller. \par
\textbf{Initialization:} Consider an arbitrary control signal $u_0$ to excite the system~(\ref{iniregulator}).
Accordingly, we define the control policy $u_a = u_0 - \Tilde{K}^k_a x_a + \Tilde{K}^k_a x_a$ with $\Tilde{K}^k_a > 0$ and feed it back to~\eqref{iniregulator} to get: 
\begin{equation} \label{sysmak}
\begin{aligned}
    \dot{x}_a&= A(0)x_a + B(0)(u_0 - \Tilde{K}^k_a x_a + \Tilde{K}^k_a x_a)\\
    & = (A(0) - B(0)\Tilde{K}^k_a)x_a + B(0)(u_0 + \Tilde{K}^k_a x_a) \\
    \dot{x}_a &= A_k(0)x_a + B(0)(u_0 + \Tilde{K}^k_a x_a).
\end{aligned}
\end{equation}
Towards eliminating the dependence of $A(0)$ and $B(0)$, we define the Lyapunov function $V^k(x_a) = x^\top_a\Tilde{P}^k_ax_a$. Taking the derivative of this function along the system~(\ref{sysmak}) leads to 
\begin{equation}
\begin{aligned}
    \frac{d}{dt}(x^\top_a\Tilde{P}_a^kx_a) =& x^\top_a(A_k^\top(0)\Tilde{P}_a^k+\Tilde{P}_a^k A_k(0)) x_a\\
    &+2 (u_0 + \Tilde{K}^k_a)^\top B(0)^\top \Tilde{P}^{k} x_a.  
\end{aligned}
\end{equation}
 Replacing $-Q_k(0) = (A_k^\top(0)\Tilde{P}_a^k+\Tilde{P}_a^k A_k(0)) = -Q(0) - \Tilde{K}^\top R(0) \Tilde{K}$ from~(\ref{ly}) and $B^\top(0)\Tilde{P}^k = R(0)\Tilde{K}^{k+1}_a$ from~(\ref{rcgain})
 leads to
\begin{equation}
\begin{aligned} \label{fin}
    \frac{d}{dt}(x^\top_a\Tilde{P}_a^kx_a) = -x^\top_a Q_k(0)x_a + 2 (u_0 + \Tilde{K}^k_a)^\top R(0)\Tilde{K}^{k+1}_a x_a 
\end{aligned}
\end{equation}
Notice that \eqref{fin} is independent of $A(0)$ and $B(0).$
Integrating both sides of~(\ref{fin}) on the interval $[t, t + \delta t]$ 
\begin{equation}
\begin{aligned}\label{pol}
    x^\top_a(t+\delta t)&\Tilde{P}_a^kx_a(t+\delta t) - x^\top_a(t)\Tilde{P}_a^kx_a(t)=\\
    &-\int_{t}^{t+\delta t} x_a^\top Q_{k}(0) x_a \; dw \\
    &2 \int_{t}^{t+\delta t}\left(\Tilde{K}^{k}_a x_a+u_{0}\right)^{T} R(0) \Tilde{K}^{k+1}_a x_a \; dw
\end{aligned}
\end{equation}
Rearranging the terms~\eqref{pol} leads to the offline policy iteration equation
\begin{equation} \label{policyiter}
\begin{aligned}
    x^\top_a(t+\delta t)&\Tilde{P}_a^kx_a(t+\delta t) - x^\top_a(t)\Tilde{P}_a^kx_a(t) \\
    & -2 \int_{t}^{t+\delta t}\left(\Tilde{K}^{k}_a x_a+u_{0}\right)^{T} R(0) \Tilde{K}^{k+1}_a x_a \; dw \\
    &= -\int_{t}^{t+\delta t} x_a^\top Q_{k}(0) x_a \; dw 
\end{aligned}
\end{equation} 
We then express~(\ref{policyiter}) in compact form using the Kronecker product ($\otimes$) 
\begin{equation} \label{lcom}
    x_a^\top \Tilde{P}^{k}_a x_a=\left(x_a^\top \otimes x_a^\top\right) \operatorname{vec}\left(\Tilde{P}^{k}_a\right),
\end{equation}
and then~\eqref{lcom} is used to transform the term from~\eqref{policyiter}
\\
$x^\top_a(t+\delta t)\Tilde{P}_a^kx_a(t+\delta t) - x^\top_a(t)\Tilde{P}_a^kx_a(t)$ 
\\
into:
\begin{equation} \label{20}
\begin{aligned}
x^\top_a(t+\delta t)\Tilde{P}_a^kx_a(t+\delta t) &- x^\top_a(t)\Tilde{P}_a^kx_a(t)= \\
    &\left[\left.x_a^\top \otimes x_a^\top\right|_{t} ^{t+\delta t}\right]\left[\operatorname{vec}(\Tilde{P}^k_a)\right].
    \end{aligned}
\end{equation}
And the next term from~\eqref{policyiter}
\begin{equation} \label{l2com}
\begin{aligned}
\left(u_{0}+\tilde{K}_{a}^{k} x\right)^{\top} & R(0) \tilde{K}_{a}^{k+1} x_{a}=\\
&\left[\left(x_{a} \top \otimes x_{a}^{\top}\right)\left(I_{n} \otimes\left(K^{k}\right)^{\top} R(0)\right)+\right.\\
&\left.\left(x_{a}^{\top} \otimes u_{0}^{\top}\right)\left(I_{n} \otimes R(0)\right)\right] \tilde{K}_{a}^{k+1}
\end{aligned}
\end{equation} 
Using~\eqref{l2com} to transform the term from~\eqref{policyiter}
\\
$-2 \int_{t}^{t+\delta t}\left(\Tilde{K}^{k}_a x_a+u_{0}\right)^{T} R(0) \Tilde{K}^{k+1}_a x_a \; dw$\\
into: 
\begin{equation} \label{21}
\begin{aligned}
 &-2 \int_{t}^{t+\delta t}\left(\Tilde{K}^{k}_a x_a+u_{0}\right)^{T} R(0) \Tilde{K}^{k+1}_a x_a \; d w = \\
 &-\left[2\left(\int_{t_{1}}^{t_{1}+\Tilde{t}} x_a^\top \otimes x_a^\top dw\right)\left(I_{n} \otimes (\Tilde{K}^{k})^\top R(0)\right)\right]\kappa  \\
 &-\left[2 \left(\int_{t_{1}}^{t_{1}+\Tilde{t}} x_a^\top \otimes u_0^\top dw\right)\left(I_{n} \otimes R(0)\right)\right]\kappa 
\end{aligned}
\end{equation}
where $\kappa = \left[\operatorname{vec}({\Tilde{K}^{k+1}_a})\right]$ and finally the last term from~\eqref{policyiter}:  
\begin{equation} \label{rcom}
    x_a^\top Q_{k}(0) x_a=\left(x_a^\top \otimes x_a^\top\right) \operatorname{vec}\left(Q_{k}(0)\right).
\end{equation}
Using~(\ref{rcom}) to transform the term $-\int_{t}^{t+\delta t} x_a^\top Q_{k}(0) x_a \; dw$ into: 
\begin{equation} \label{22}
\begin{aligned}
-\int_{t}^{t+\delta t} &x_a^\top Q_{k}(0) x_a \; dw =\\
&-\left[\int_{t_{1}}^{t_{1}+\delta {t}} x_a^\top \otimes x_a^\top dw\right]\operatorname{vec}\left(Q_k(0)\right)
\end{aligned}
\end{equation}
Using~(\ref{20}),~(\ref{21}) and~(\ref{22}) to express the offline policy iteration~(\ref{policyiter}) in a compact form:
\begin{equation}
\begin{aligned}
 &\left[\left.x_a^\top \otimes x_a^\top\right|_{t} ^{t+\delta t}\right]\left[\operatorname{vec}(\Tilde{P}^k_a)\right] \\ &-2\left[\left(\int_{t_{1}}^{t_{1}+\Tilde{t}} x_a^\top \otimes x_a^\top dw\right)\left(I_{n} \otimes (\Tilde{K}^{k})^\top R(0)\right)\right]\kappa\\
 & -2\left[\left(\int_{t_{1}}^{t_{1}+\Tilde{t}} x_a^\top \otimes u_0^\top dw\right)\left(I_{n} \otimes R(0)\right)\right]\kappa \\
 & = -\left[\left.x_a^\top \otimes x_a^\top\right|_{t} ^{t+\delta t}\right]\operatorname{vec}\left(Q_k(0)\right)
\end{aligned}\label{compact1}
\end{equation}
We can now express \eqref{compact1} in the compact matrix form as 
\begin{equation} \label{compact}
        \Tilde{\psi}\left[\begin{array}{c}
\operatorname{vec}\left(\Tilde{P}^{k}_a\right) \\
\operatorname{vec}\left(\Tilde{K}^{k+1}_a\right)
\end{array}\right]=\Tilde{\Gamma}
\end{equation}
where, \\
$\tilde{\psi}=\left[\tilde{\delta}_{x x},-2 \tilde{I}_{x x}\left(I_{n} \otimes\left(\tilde{K}^{k}\right)^{\top} R(0)\right)-2 \tilde{I}_{x \bar{u}_{0}}\left(I_{n} \otimes R\right)\right]$, 
$\Tilde{\delta}_{x x} = \left[\left.x_a^\top \otimes x_a^\top\right|_{t} ^{t+\delta t}\right]$, \\
$\Tilde{I}_{x x} = -2\left[\left(\int_{t_{1}}^{t_{1}+\Tilde{t}} x_a^\top \otimes x_a^\top dw\right)\left(I_{n} \otimes (\Tilde{K}^{k})^\top R(0)\right)\right]$ and \\
$\Tilde{I}_{x u_0} = -2\left[\left(\int_{t_{1}}^{t_{1}+\Tilde{t}} x_a^\top \otimes u_0^\top dw\right)\left(I_{n} \otimes R(0)\right)\right]$.

This way offline policy iteration~(\ref{policyiter}) is expressed in compact form~(\ref{compact}). It should be emphasized that the learning is going to be done through the use of the state $x(t)$ according to Remark~(\ref{remark1}). In the next step, we collect this data offline to estimate the feedback control gain.  \par
\textbf{Data Collection:} During learning, we collect online measurement data, including state space $x(t)$ and control policy $u_0$, for the time period $[t_i, t_j]$ with the sampling interval $t_{i+1} - t_i = \delta t = \Tilde{t}$. This is followed by computing the matrices $\delta_{x x}$, $I_{x x}$ and $I_{x u_0}$ as follows:
\begin{equation} 
\delta_{x x}=\left[\left.x^\top \otimes x^\top\right|_{t_{1}} ^{t_{1}+\Tilde{t}}, \quad \ldots,\left.\quad x^\top \otimes x^\top\right|_{t_{j}} ^{t_{j}+\Tilde{t}}\right]^\top
\end{equation}
\begin{equation}
I_{x x}=\left[\int_{t_{1}}^{t_{1}+\Tilde{t}} x^\top \otimes x^\top dw,  \ldots, \quad \int_{t_{l}}^{t_{l}+\Tilde{t}} x^\top \otimes x^\top d w\right]^\top
\end{equation}
\begin{equation}
I_{x u_0}=\left[\int_{t_{1}}^{t_{1}+\Tilde{t}} x^\top \otimes u_0^\top dw,  \ldots, \quad \int_{t_{l}}^{t_{l}+\Tilde{t}} x^\top \otimes u_0^\top d w\right]^\top
\end{equation} \par
\begin{assumption}\label{as3}
There exists a large number $j > 0$ such that rank([$I_{x x} \quad I_{x u_0}$]) = $\frac{n(n+1)}{2} + mn$.
\end{assumption}
Assumption \ref{as3} ensures the collection of enough data for the learning process~\cite{c26}. 

\textbf{Policy Iteration}: This step further involves two sub-steps. i) Policy evaluation; Evaluating the effectiveness of the current policy using the data from the present data step.  and ii) Policy improvement; After we evaluate the current policy we then update the past policy with this new one. These two steps will be performed simultaneously until the threshold is met. 
This is given as: 
\begin{equation} \label{iterationeq1}
\Tilde{\psi}_{k}\left[\begin{array}{c}
\operatorname{vec}\left(\Tilde{P}^{k}_a\right) \\
\operatorname{vec}\left(\Tilde{K}^{k+1}_a\right)
\end{array}\right]=\Tilde{\Gamma}_{k},
\end{equation}
where
%\vspace{-2 mm}
\begin{equation} \label{compacted}
\begin{aligned}
\tilde{\psi}_{k}=\left[\delta_{x x},-2 I_{x x}\right.&\left(I_{n} \otimes\left(\tilde{K}^{k}\right)^{\top} R(0)\right) \\
&\left.-2 I_{x \tilde{u}_{0}}\left(I_{n} \otimes R\right)\right],
\end{aligned}
\end{equation}
%\vspace{-12 mm}
\begin{equation} \label{compacted2}
\Tilde{\Gamma}_{k}=-I_{x x} \operatorname{vec}\left(Q_{k}\right).
\end{equation}
%\vspace{-2 mm}

At the end of the learning (i.e at the end of convergence), the feedback gain can be obtained as $K_a = \Tilde{K}^{k+1}_a$ and the controller for the system~(\ref{iniregulator}) is $u_a = -K_ax_a$.

(ii) Terminal Regulator problem: We follow the same steps as in the initial regulator problem except the initialization of the control gain should satisfy $K_b < 0$. Note that controller for the system~(\ref{finregulator}) is $u_b = -K_b^*x_b.$ Where, $K_b^*$ is the optimal feedback gain control. 

The pseduocode for the learning algorithm is given in Algorithm~(1)
\begin{algorithm}[h] \label{algo}
\SetAlgoLined
Initialize $K_a > 0$ \;
\While{$rank([I_{x x} \quad I_{x u_0}]) < \frac{n(n+1)}{2} + mn $}{
  Collect the online measurement data $x(t)$ using the excitation control $u_a = u_0$\;
Construct the matrices $\delta_{x x}$, $I_{x x}$ and $I_{x u_0}$}
 \While{$|P^k_a - P^{k+1}_a | <$ Threshold}{
  Estimate the values of $P_a^k$ and $K_a^{k+1}$ through \eqref{iterationeq1}
 }
$u_a = -K_a^{k+1}x$\;
%\quad \forall t \in [0, T] $\\
Initialize $K_b < 0$\;
\While{$rank([I_{x x} \quad I_{x u_0}]) < \frac{n(n+1)}{2} + mn $}{
  Collect the online measurement data $x(t)$ using the excitation control $u_b = u_0.$\\
Construct the matrices $\delta_{x x}$, $I_{x x}$ and $I_{x u_0}$
 }
\While{$|P^k_b - P^{k+1}_b | <$ Threshold}{
  Estimate the values of $P_b^k$ and $K_b^{k+1}$ through \eqref{iterationeq1}
 }
$u_b = -K_b^{k+1}x$
\caption{Offline policy iteration on two value boundary problems}
\end{algorithm}
\addtolength{\textheight}{-3cm}
$\newline$

Following the learning procedure described in this section, and in a similar way to the result reported in Theorem \ref{theorem}, the closed-loop performance of the learned-control system, comprised of \eqref{systemeq} and $u_\text{learned}=u_a+u_b,$ will approximate that of the original control system, comprised of \eqref{systemeq} and \eqref{control}, provided that $\varepsilon$ is sufficiently small or $T$ is sufficiently long. This will be verified in the simulation example given in the next section.  

\section{Simulation Example} \label{simulation}
\begin{figure*}[ht] 
    \centering
    \subfloat[\centering state trajectory for $\varepsilon$ = 0.9]{\includegraphics[width=5.45cm]{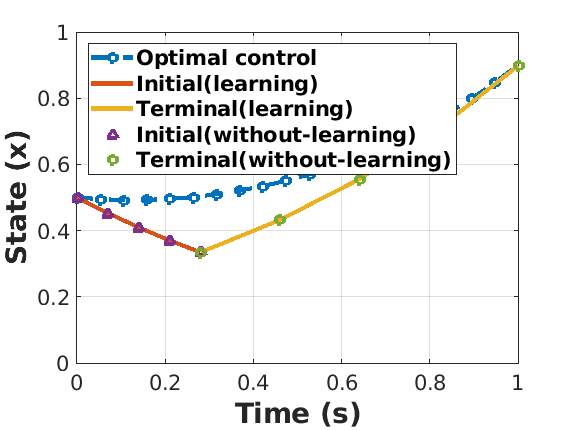}}%
    \qquad
    \subfloat[\centering state trajectory for $\varepsilon$ = 0.5]{{\includegraphics[width=5.45cm]{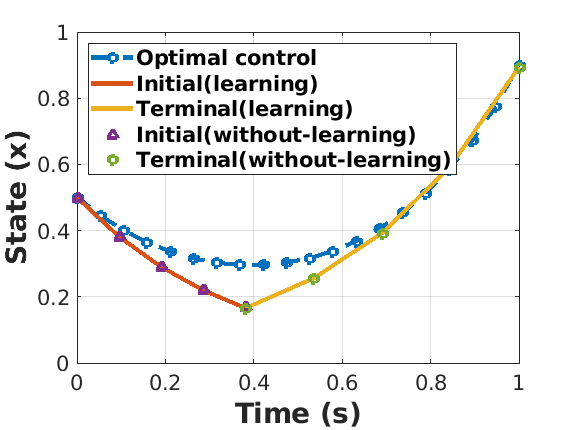}}}%
    \qquad
    \subfloat[\centering state trajectory for $\varepsilon$ = 0.1]{{\includegraphics[width=5.45cm]{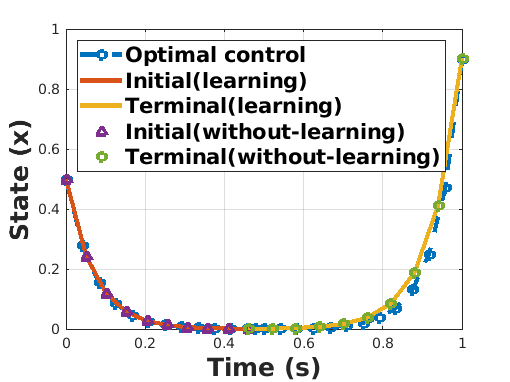}}}%
    \qquad
    \vskip 0.2in
    \subfloat[\centering Controller for $\varepsilon$ = 0.9]{\includegraphics[width=5.45cm]{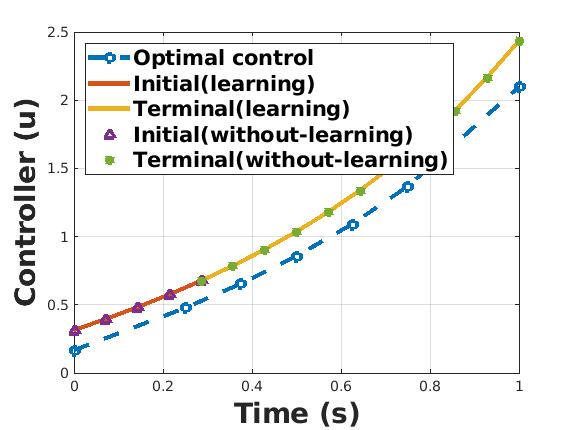}}%
    \qquad
    \subfloat[\centering Controller for $\varepsilon$ = 0.5]{{\includegraphics[width=5.45cm]{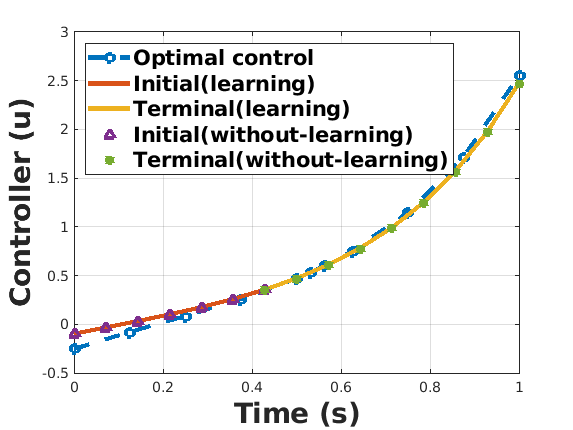}}}%
    \qquad
    \subfloat[\centering Controller for $\varepsilon$ = 0.1]{{\includegraphics[width=5.45cm]{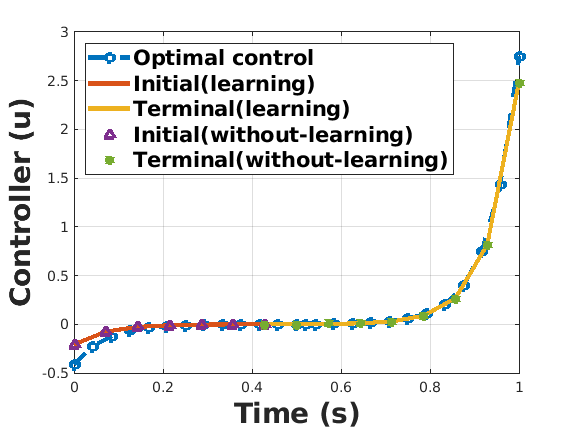}}}%
    \caption{Plots (a), (b) and (c) illustrate the state space trajectory for the state $x(t)$ for different values of $\varepsilon$ and Plots (d), (e) and (f) illustrates the control law $u(t)$ for different values of $\varepsilon$.}
    \label{resultsplot}
\end{figure*}
Consider an object with a dissipating mass $m(t)$ on a frictionless environment that is described by the following equation \cite{c101}
\begin{equation} \label{objectdynami}
    \frac{dx}{dt} = \frac{1}{m(t)}x + B(t)u
\end{equation}
where $x(t)$ is the position of the body. We assume that the time-varying dissipating mass is unknown.
%footnote{This problem is solved in \cite{c101} and \cite{c102} assuming that the mass model is known.}
In the absence of the external disturbances, the force $u(t)$ is applied on the object to move it from one position to another.
A controller is sought to achieve the control objective while keeping the force and position minimum in between the initial and final values. That is, the controller is needed to minimize the objective function 
\begin{equation} \label{obfun}
    J = \frac{1}{2}\int_{0}^{T} q(t)x^2(t) + r(t)u^2(t) \hspace{2mm} dt
\end{equation}
The initial and final conditions of the system are specified as: $x_0 = 0.5$ and $x_T = 0.9$ and the cost function parameters are specified as $q(t)=1$, $r(t)=1$.
For simulation purposes, we assume that the mass is given by $m(t) = \frac{-1}{(1+0.2t)},$ and $B(t)=1.$ During the offline data collection, we excite the system using the controller $u_0 = \sum_0^{100} sin(wt)$ with the time interval of 0.01 seconds, where, $w$ is the frequency. where, $w$ is the random value. The control gains $K_a$ and $K_b$ are found using the learning method described in Section \ref{learning} and found to be 0.410 and -2.46, respectively. 

To compare our results, we found $K_a$ and $K_b$ values assuming that $A(t)$ and $B(t)$ are known through the procedure outlined in Section \ref{Main result} by solving the problems \eqref{iniregulator}-\eqref{j_b}. Accordingly, the Riccati equation for the system~\eqref{objectdynami} for which the controller needs to minimize the objective function~\eqref{obfun} is
\begin{equation} \label{ex}
\begin{aligned}
%-(1+0.2t)^\top P- P (1+0.2t) - P^2 + 1 &= 0 \\
%-2(1+0.2t)P - P^2 + 1 &= 0\\
P^2 + 2(1+0.2t)-1 &= 0.
\end{aligned}
\end{equation}
The two roots for the Riccati equation~\eqref{ex} are:
\begin{equation}
    P = -(1+0.2t) \pm \sqrt{(1+0.2)^2 + 1}.
\end{equation}
Since $P_a \geq 0$ and $P_b \leq 0$ the values of $P_a(0)$ and $P_b(1)$ are: 
\begin{equation}
    P_a(0) = -1+\sqrt{2}=0.414, \quad P_b(1) = -1 - \sqrt{2}=-2.414.
\end{equation}

After finding the controllers $u_a$ and $u_b$, we simulate the system through finding the time $t_c$ for which $x_a(t_c) = x_b(1-t_c)$. Recall that $x_a$ and $x_b$ refer to the state trajectories in the forward and reverse time respectively. We then use the controller $u_a$ in the time interval $[0, t_c]$ and the controller $u_b$ in the time interval $[t_c, 1]$. Note that we are normalizing the time interval to $[0,1]$ instead of $[0,T]$ to make the comparison between the controllers clear. 

% we follow two steps to simulate the state space trajectory of the system.
% \begin{enumerate}
%     \item Find the time $t_c$ for which $x_a(t_c) = x_b(1-t_c)$
%     \item Use the controller $u_a = -K_ax$ on system~\eqref{iniregulator} in the time interval $[0, t_c]$ and the controller $u_b = -K_ax_b$ on the system~\eqref{finregulator} in the time interval $[t_c, 1]$
% \end{enumerate}
% The systems are then controlled in the given intervals.
% \begin{equation}
% \begin{aligned}
% \varepsilon \dot{x}_a &= A(t)x_a + B(t)u_a, \qquad0 \leqslant t \leqslant t_{\mathrm{c}}, \qquad x_a(0) = x_0,\\
% \varepsilon \dot{x}_b &= A(t)x_b + B(t)u_b, \qquad t_{\mathrm{c}}<t \leqslant 1, \qquad x_b(t_c) = x_c,
% \end{aligned}
% \end{equation}
% where $x_c$ is the value of the state space $x_a(t_c)$ at the time $t_c$, which is the end point for the system~\eqref{iniregulator} when using the controller $u_a = -K_ax_a$. The end point value now would be the initial condition for the system~\eqref{finregulator} when the controller is $u_b = -K_bx_b$.
% \addtolength{\textheight}{-3cm}

 The optimal state trajectory and the optimal controller is found using the two value boundary problem solver from MATLAB command \textit{bvp4c} using the knowledge of $A(t)$ and $B(t)$. We then compare the state space trajectory and the controller obtained using our learning method with that of the approximate method and the optimal controller with $A(t)$ and $B(t)$ known.
From Figure~\ref{resultsplot}, we can see that the learning-based controller accurately approximates the optimal controller as $\varepsilon$ is decreased (or control time period $T$ increases). This implies that the performance of the learning controller is a sub-optimal one and converges to the optimal performance as the time interval $T$ gets large. 

\section{CONCLUSIONS AND FUTURE WORKS}
We proposed optimal controller design using reinforcement learning for time varying systems with bounded conditions. The proposed design leverages the fast time scale occurring at the boundary conditions to reduce the time-varying problem into two simple time-invariant problems. Further, we design the sub-optimal controller to approximate the controller for time-varying systems. The accuracy of the controller performance improves as the problem time horizon increases. We presented simulation results to support our claims using a dissipating mass system on a friction-less environment. In the future, we plan to extend our work in learning the controllers of nonlinear systems.

\addtolength{\textheight}{-3cm}

\end{document}